\documentclass[aps,prb,twocolumn,groupedaddress,showpacs]{revtex4-1}

\usepackage{color}
\usepackage{graphicx}
\usepackage{dcolumn}
\usepackage{bm}
\usepackage{float}
\usepackage[mathlines]{lineno}

\usepackage{amsmath}

\begin{document}

\title{Pair-checkerboard antiferromagnetic order in $\beta$-Fe$_4$Se$_5$ with $\sqrt{5}\times\sqrt{5}$ ordered Fe vacancies}

\author{Miao Gao$^{1}$}
\author{Xin Kong$^2$}
\author{Xun-Wang Yan$^{3}$}
\author{Zhong-Yi Lu$^{4}$}\email{zlu@ruc.edu.cn}
\author{Tao Xiang$^{2,5}$}\email{txiang@iphy.ac.cn}

\affiliation{$^{1}$Department of Microelectronics Science and Engineering, Faculty of Science, Ningbo University, Zhejiang 315211, China}

\affiliation{$^{2}$Institute of Physics, Chinese Academy of Sciences, Beijing 100190, China}

\affiliation{$^{3}$School of physics and electrical engineering, Anyang Normal University, Henan 455000, China}

\affiliation{$^{4}$Department of Physics, Renmin University of China, Beijing 100872, China}

\affiliation{$^{5}$Collaborative Innovation Center of Quantum Matter, Beijing, China}

\date{\today}

\begin{abstract}
  The electronic structure of recently discovered $\beta$-Fe$_4$Se$_5$ with $\sqrt{5} \times \sqrt{5}$ ordered Fe vacancies is calculated using  first-principles density functional theory.
  We find that the ground state is an antiferromagnetic (AFM) insulator in agreement with the experimental observation.
  In K$_2$Fe$_4$Se$_5$, it is known that the ground state is $\sqrt{5} \times \sqrt{5}$-blocked-checkerboard AFM ordered.
  But for this material, we find that the ground state is $\sqrt{5} \times \sqrt{5}$-pair-checkerboard AFM ordered, in which the intrablock four Fe spins exhibit the collinear AFM order and the interblock spins on any two nearest-neighboring sites are antiparallel aligned.
  This state is about 130 meV/Fe lower in energy than the $\sqrt{5} \times \sqrt{5}$-blocked-checkerboard AFM one.
  Electron doping can lower the energy of the $\sqrt{5} \times \sqrt{5}$-blocked-checkerboard AFM state and introduce a transition between these two states, suggesting that there is strong AFM fluctuation in FeSe-based materials upon doping.
  This provides a unified picture to understand the AFM orders in $\beta$-Fe$_4$Se$_5$ and in alkali-metal intercalated FeSe materials.
\end{abstract}


\maketitle

\section{Introduction}

The discovery of iron-based superconductors has opened a new path to explore unconventional mechanism of high-temperature superconductivity \cite{Kamihara-JACS}. The superconducting transition temperatures of iron-based superconductors are closely related to the height of anion from the Fe-Fe square lattice and the Fe-(As/Se)-Fe angle \cite{Zhao-NatMater7,Okabe-PRB,Mizuguchi-SST}.
By intercalating atoms between FeSe layers to form $A_{1-x}$Fe$_{2-y}$Se$_2$ compounds ($A$=K, Tl, Rb, or Cs), one can effectively modulate the height of anion and raise the superconducting transition temperature of bulk FeSe from 8 K~ \cite{Hsu-PNAS} to 27$\sim$48 K~ \cite{Guo-PRB82,Maziopa-JPCM,Fang-EPL,Wang-EPL93,Sun-Nature,Ying-Sci2}.

$A_{1-x}$Fe$_{2-y}$Se$_2$ possesses different kinds of Fe-vacancy orders. For example, the two commonly studied K$_{2x}$Fe$_{2-x}$Se$_2$ compounds,  K$_2$Fe$_3$Se$_4$ (KFe$_{1.5}$Se$_2$) and K$_2$Fe$_4$Se$_5$ (K$_{0.8}$Fe$_{1.6}$Se$_2$), have the $\sqrt{2}\times2\sqrt{2}$ and $\sqrt{5}\times\sqrt{5}$ Fe-vacancy orders, respectively.
These two compounds were predicted to be AFM semiconductors by the first-principles calculations~\cite{Yan-PRL,Yan-PRB83}, and later confirmed by the experimental measurements~\cite{Zhao-PRL109,Bao-CPL28,Chen-PRX1}.
Other kind of Fe-vacancy orders has also been observed~\cite{Ding-Natcomm4}.
The stoichiometric compound KFe$_2$Se$_2$ without Fe vacancies is unstable against mesoscopic phase separation~\cite{Wang-PRB83,Pomjakushin-JPCM,Ricci-PRB84,Ricci-PRB91}.
But it can be stablized and become a high-T$_c$ superconductor~\cite{Li-PRL109,Zhang-PRB} by proximity to K$_2$Fe$_4$Se$_5$.

Recently, a series of $\beta$-Fe$_{1-x}$Se materials without metal intercalations but with Fe vacancies were synthesized.
Through the measurement of electron microscopy, Fe vacancies are also found to be ordered and mainly in the phases with $\sqrt{2}\times\sqrt{2}$, $\sqrt{5}\times\sqrt{5}$, and $\sqrt{10}\times\sqrt{10}$ vacancy orders~\cite{Chen-PNAS}.
It indicates that the Fe-vacancy order is a common feature of FeSe-based superconductors.
Furthermore, it was found that most of $\beta$-Fe$_{1-x}$Se samples synthesized in laboratory are in the $\sqrt{5}\times\sqrt{5}$ phase, and the compound with such a vacancy order, i.e. $\beta$-Fe$_4$Se$_5$, is an AFM insulator \cite{Chen-PNAS,Fang-PRB93}.
As both $\beta$-Fe$_4$Se$_5$ and K$_2$Fe$_4$Se$_5$ have the same Fe-vacancy orders, it is interesting to know whether they have the same magnetic order in the ground states.

In this paper, we present the result for the electronic and magnetic structures of $\beta$-Fe$_4$Se$_5$ with the $\sqrt{5}\times\sqrt{5}$ Fe-vacancy order obtained by the first-principles density functional calculations.
We find that $\beta$-Fe$_4$Se$_5$ is an insulator with a $\sqrt{5} \times \sqrt{5}$-pair-checkerboard AFM order.
In this AFM ordered state, each Fe block contains four Fe spins which are collinear AFM aligned.
The band gap of $\beta$-Fe$_4$Se$_5$ in this $\sqrt{5}\times\sqrt{5}$-pair-checkerboard AFM phase is found to be 290 meV, in agreement with the experimental observation \cite{Chen-PNAS,Fang-PRB93}.
This AFM order is stable against light electron doping.
It is replaced by the $\sqrt{5}\times\sqrt{5}$-blocked-checkerboard AFM order (which is also the AFM order observed in K$_2$Fe$_4$Se$_5$) when the doping level excesses 1.7 electrons per Fe$_4$Se$_5$ formula.

\section{COMPUTATIONAL DETAILS}

In our calculations the plane wave basis method is used~\cite{pwscf}.
We adopt the generalized gradient approximation (GGA) of Perdew-Burke-Ernzerhof \cite{pbe} for the exchange-correlation potentials.
The ultrasoft pseudopotentials \cite{vanderbilt} are used to model the electron-ion interactions.
After full convergence test, the kinetic energy cut-off and the charge density cut-off of the plane wave bases are taken to be 45 Ry and 480 Ry, respectively.
Our calculation is done with a $\sqrt{5}\times\sqrt{5}\times1$ tetragonal unit
cell which contains one FeSe layer with 8 Fe atoms, 2 vacancies, and 10 Se atoms.
A mesh of 8$\times$8$\times$10 {\bf k}-points is used to sample the
Brillouin zone of the $\sqrt{5}\times\sqrt{5}\times1$ unit cell.
The Gaussian broadening technique of width 0.002 Ry is used in the calculation of metallic states.
The convergence is tested with a denser 16$\times$16$\times$20 {\bf k}-mesh.
There are two degenerate ground states, corresponding to the two degenerate chiral structures of the $\sqrt{5}\times\sqrt{5}$ lattice \cite{Yan-PRB83,Sabrowsky-JMMM}.
We choose the right-chiral structure to do the calculation.

\section{RESULTS AND ANALYSIS}

\begin{figure}[t]
\begin{center}
\includegraphics[width=8.6cm]{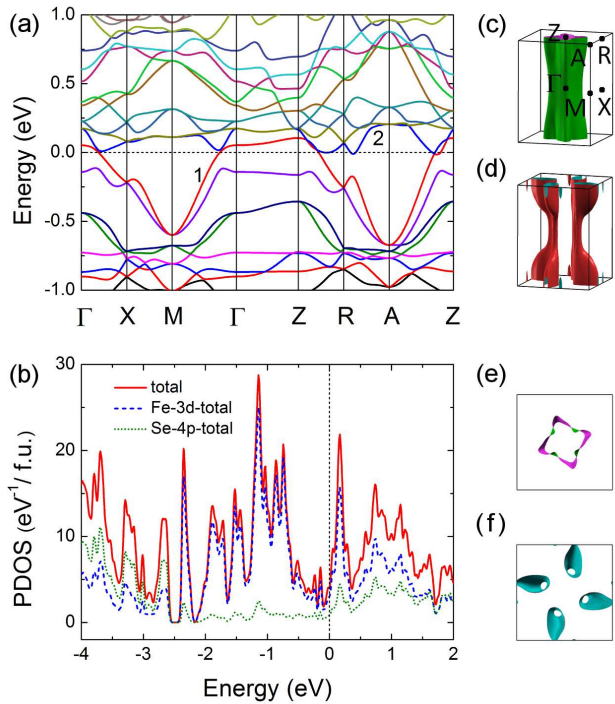}
  \caption{(Color online) Electronic structures of $\beta$-Fe$_4$Se$_5$ with the  $\sqrt{5}\times\sqrt{5}$ Fe-vacancy order in the nonmagnetic phase.
  (a) Band structures. The Fermi energy is set to zero.
  (b) Orbital-resolved partial density of states per formula of $\beta$-Fe$_4$Se$_5$.
  (c) and (d) Fermi surfaces corresponding to the energy bands labelled by the Arabic numbers, 1 and 2, shown in (a), respectively. The high-symmetry points in the Brillouin zone are given in (c).
  (e) and (f) Top views of the two Fermi surfaces.}
\label{fig:NM-Band}
\end{center}
\end{figure}

\begin{figure}[H]
\begin{center}
\includegraphics[width=8.6cm]{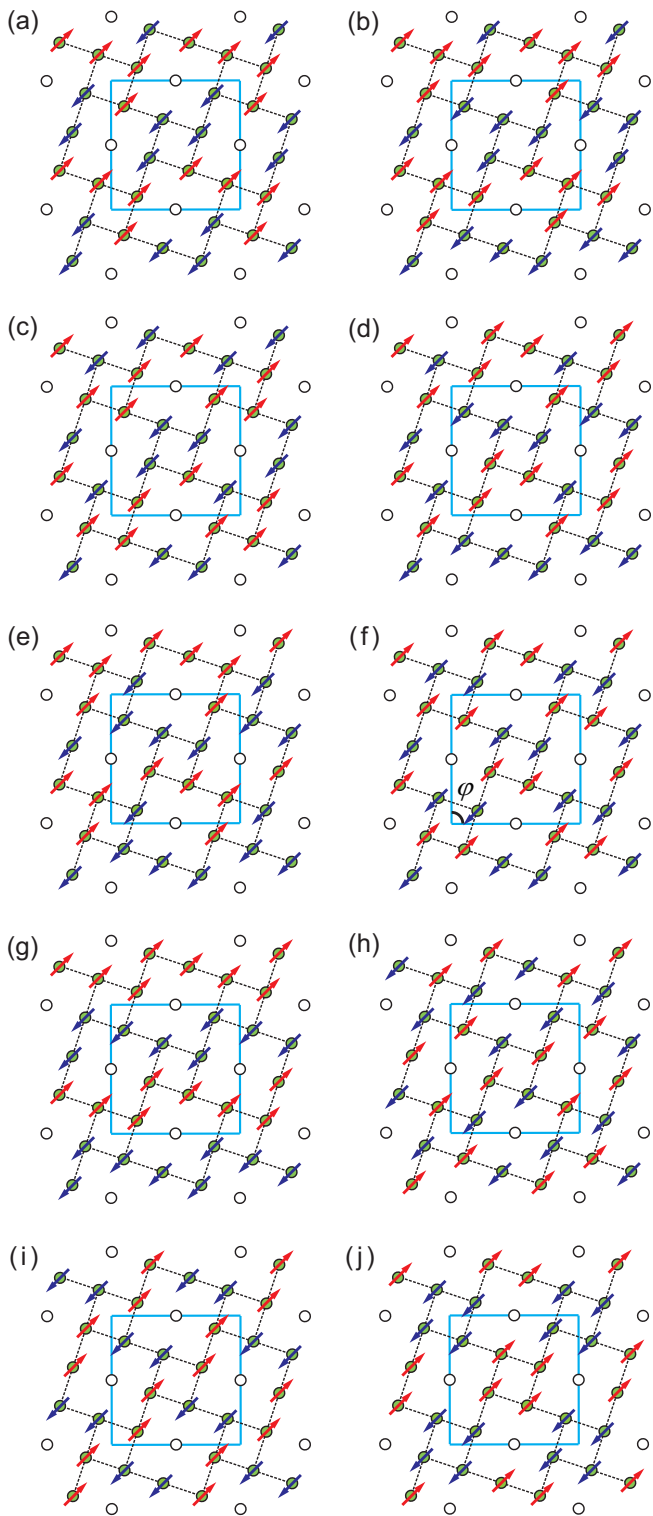}
  \caption{(Color online) Schematic top views of ten possible AFM orders, labelled from AFM-$a$ to AFM-$j$, in each FeSe layer.
  The $\sqrt{5}\times\sqrt{5}$-pair-checkerboard AFM order shown in (f) is the magnetic structure for the ground-state of $\beta$-Fe$_4$Se$_5$.
  The squares enclosed by the blue lines denote the unit cells.
  The filled and empty circles represent Fe atoms and Fe vacancies, respectively.
  The up and down Fe spins are represented by red and blue arrows, respectively.}
\label{fig:r5xr5}
\end{center}
\end{figure}

For the vacancy ordered $\beta$-Fe$_4$Se$_5$, there are eight Fe atoms in
each $\sqrt{5}\times \sqrt{5} \times1$ unit cell. These eight Fe atoms can be divided into two blocks. One block contains the four Fe atoms in the middle of the unit cell enclosed by the blue box as shown in Fig.~\ref{fig:r5xr5}, and the other contains the rest four Fe atoms.

The nonmagnetic (NM) phase of $\beta$-Fe$_4$Se$_5$ is metallic.
Figure \ref{fig:NM-Band} shows the band structure and the partial density of states in the NM state.
There are two bands crossing the Fermi level, forming one pillar-shaped hole Fermi surface along the $\Gamma$-Z line [Figs.~\ref{fig:NM-Band}(c) and \ref{fig:NM-Band}(e)] and one windmill-type electron Fermi surfaces around the zone center [Figs.~\ref{fig:NM-Band}(d) and \ref{fig:NM-Band}(f)].
The low-energy physics is governed by Fe $3d$ orbitals, since the density of states (DOS) in the energy range from -2.5 eV to 1.0 eV is contributed mainly by Fe $3d$ electrons.

\begin{table}
  \caption{Total energies, averaged magnitudes of the magnetic moments of Fe spins ($|M|$), and band gaps of the ten AFM orders shown in Fig.~\ref{fig:r5xr5}.
  The energy of NM is set to zero.}
  \label{table:energy}
\begin{tabular}{cccc}
  \hline
  \hline
  AFM states & energy &  \,\, $|M|$ \,\, & \,\, band gap \,\,\\
  & \,\, (meV/Fe) \,\, & ($\mu_\text{B}$) &  (meV) \\
  \cline{1-4}
  AFM-$a$ & -127.4 & 2.48 & 0.0  \\
  AFM-$b$ & -77.8 & 2.53 & 82.6  \\
  AFM-$c$ & -188.9 & 2.72 & 208.5  \\
  AFM-$d$ & -155.5 & 2.66 & 163.6  \\
  AFM-$e$ & -114.0 & 2.53 & 75.7  \\
  AFM-$f$ & -245.6 & 2.70 & 290.0  \\
  AFM-$g$ & -20.3 & 2.29 & 0.0  \\
  AFM-$h$ & -168.4 & 2.47 & 0.0  \\
  AFM-$i$ & -207.7 & 2.66 & 20.8  \\
  AFM-$j$ & -115.8 & 2.35 & 0.0  \\
  \hline
  \hline
\end{tabular}
\end{table}

The ground state, or the lowest energy state, of $\beta$-Fe$_4$Se$_5$ in the $\sqrt{5}\times\sqrt{5}$ vacancy ordered phase is AFM ordered.
In order to determine the magnetic structure of the ground state, we have calculated the electronic structures for all ten non-equivalent AFM states allowed in a $\sqrt{5}\times\sqrt{5}$ lattice.
Figure \ref{fig:r5xr5} shows the spin configurations of these ten AFM states, labelled from AFM-$a$ to AFM-$j$.
These states can be divided into three groups, depending on the number of up spins in the middle four Fe atoms, which is called an Fe block, within each unit cell enclosed by the blue square.
The first group contains the spin configurations represented by (AFM-$a$)-(AFM-$c$) with just one up spin in an Fe block.
The second one consists of the spin configurations represented by (AFM-$d$)-(AFM-$i$) with two up spins in an Fe block.
There is one spin configuration, i.e. AFM-$j$, in the third group.
It contains four up spins in an Fe block.
The AFM states with one and three up Fe spins in an Fe block are energetically degenerate and physically equivalent.
The states with two up spins inside an Fe block can have either a collinear AFM order [Figs.~\ref{fig:r5xr5}(d)-(g)] or a checkerboard one [Figs.~\ref{fig:r5xr5}(h) and \ref{fig:r5xr5}(i)].
Among these states, AFM-$j$ is the $\sqrt{5}\times\sqrt{5}$-blocked-checkerboard AFM order, which is the magnetic order observed in the ground state of K$_2$Fe$_4$Se$_5$.

The relative energies of the ten AFM states with respect to the energy of the NM state are given in Table I.
The ground state of $\beta$-Fe$_4$Se$_5$ is that shown in Fig.~\ref{fig:r5xr5}(f), namely the state labelled by AFM-$f$.
It is a $\sqrt{5}\times\sqrt{5}$-pair-checkerboard AFM state.
A similar AFM state, called the pair-checkerboard AFM state, was found to be the ground state of bulk $\beta$-FeSe based on the first principles

\noindent electronic structure calculations \cite{Cao-arXiv1407}.
AFM-$f$ can be regarded as a realization of this pair-checkerboard AFM order on the $\sqrt{5}\times\sqrt{5}$ Fe-vacancy ordered lattice.

In the ground state, the lattice undergoes a tetragonal to monoclinic structure transition induced by the spin-lattice interaction.
This changes the angle, $\varphi$, defined in Fig.~\ref{fig:r5xr5}(f) from 90$^\circ$ to 89.2$^\circ$ with an energy gain of 0.8 meV/Fe.
Consequently, there is a tiny difference between the lattice parameters along the $a$ and $b$ axes, which are equal to 8.4519 {\AA} and 8.4855 {\AA}, respectively.
This kind of structure transition is similar to the tetragonal-orthorhombic phase transition observed in other parent compounds of iron-based superconductors \cite{Cruz-Nature,Rotter-PRB78,McQueen-PRL103,Ma-PRL102}.

\begin{figure}[t]
\includegraphics[width=8.6cm]{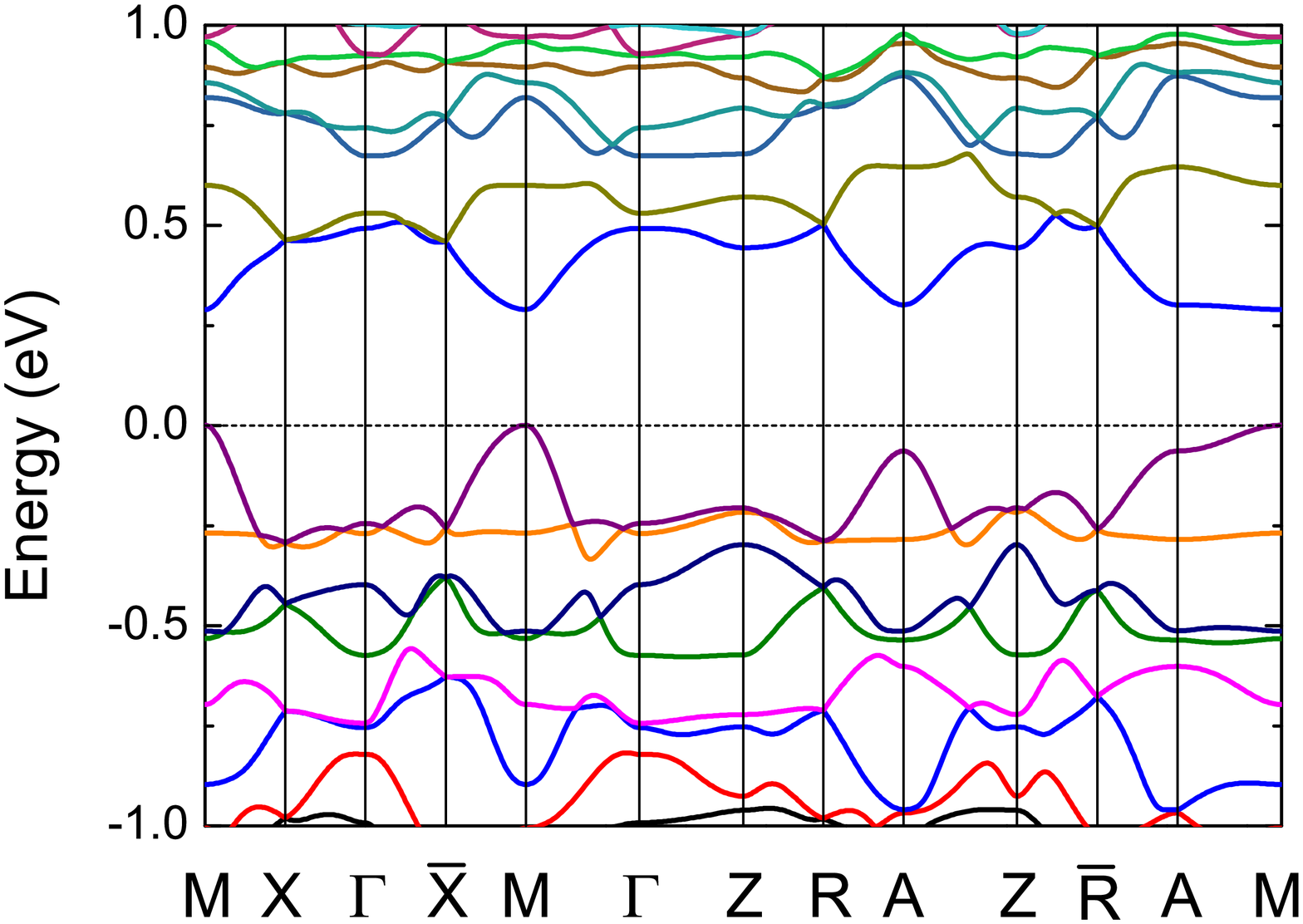}
  \caption{(Color online) Calculated electronic band structure of $\beta$-Fe$_4$Se$_5$ in the $\sqrt{5}\times\sqrt{5}$-pair-checkerboard AFM state with a tetragonal unit cell.
  The top of the valance band energy is set to zero.
  The Brillouin zone is the same as that in Fig.~\ref{fig:NM-Band}(c). The fractional coordinates of $X$, $\bar{X}$, $R$ and $\bar{R}$ with respect to the reciprocal lattice vectors are (0.0, 0.5, 0.0), (0.5, 0.0, 0.0), (0.0, 0.5, 0.5) and (0.5, 0.0, 0.5), respectively. }
\label{fig:Col-Band}
\end{figure}

Figure \ref{fig:Col-Band} shows the band structure of $\beta$-Fe$_4$Se$_5$ in the $\sqrt{5}\times\sqrt{5}$-pair-checkerboard AFM state obtained with a tetragonal unit cell by ignoring the monoclinic distortion.
It indicates clearly that $\beta$-Fe$_4$Se$_5$ is an AFM insulator with a direct energy gap of about 290 meV, in agreement with the experimental observation \cite{Chen-PNAS,Fang-PRB93}.

\begin{figure}[t]
\includegraphics[width=8.6cm]{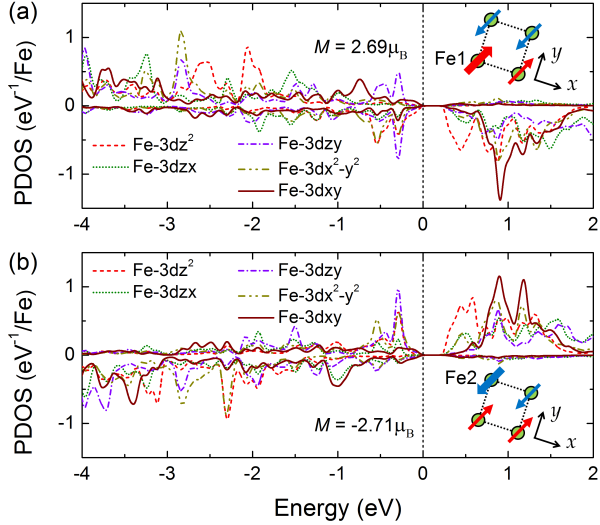}
  \caption{(Color online) Orbital-resolved partial density of states (PDOS) for the five Fe $3d$ orbitals at (a) site Fe$_1$ and (b) site Fe$_2$ in the $\sqrt{5} \times \sqrt{5}$-pair-checkerboard AFM state.
  The spins of Fe$_1$ and Fe$_2$ are highlighted by large arrows. The magnetic moments of these two Fe atoms are given in the corresponding figures. }
\label{fig:Col-PDOS}
\end{figure}

Figure \ref{fig:Col-PDOS} shows the orbital-resolved partial DOS in the $\sqrt{5}\times\sqrt{5}$-pair-checkerboard AFM state with tetragonal symmetry.
This AFM state is invariant under the following two kinds of transformations:
One is to combine an $ab$-plane reflection with a fractional translation of (-1/2, 1/2, 0), the other is to take a spin inversion and then a 180$^\circ$ rotation around the $c$-axis.
This symmetry property ensures the spin-up/down DOS at an Fe atom to equal the spin-down/up DOS at the diagonal Fe atom inside the same block.
Thus we need only to show the orbital-resolved partial DOS for two Fe atoms, labelled by Fe$_1$ and Fe$_2$ in the figure.
For Fe$_1$, the five up-spin orbitals are almost completely filled, but the five down-spin orbitals are only partially filled.
The DOS at Fe$_2$ behaves similarly, but the up and down spins are reversed.
The magnetic moment of Fe$_2$ is slightly larger than that of Fe$_1$.
This leads to the difference in the DOSs between the up/down-spin part of Fe$_1$ and the down/up-spin one of Fe$_2$.
The unoccupied states are mainly contributed by $d_{z^2}$ and $d_{xy}$ orbitals.

We also find that the AFM states defined by AFM-$b$, AFM-$c$, AFM-$d$, AFM-$e$, and AFM-$i$ are insulating.
The excitation energy gaps are equal to 82.6 meV, 208.5 meV, 163.6 meV, 75.7 meV, and 20.8 meV, respectively.
For AFM-$d$, the band gaps for the up- and down-spin electrons are not equal to each other, since the corresponding up- and down-spin configurations are not equivalent.
They equal 163.6 meV and 233.1 meV, respectively.
The other AFM states, including the $\sqrt{5}\times\sqrt{5}$-blocked-checkerboard AFM state, i.e. AFM-$j$, are metallic.
This differs from the case in K$_2$Fe$_4$Se$_5$ \cite{Yan-PRL,Bao-CPL28,Chen-PRX1} where
the ground state is blocked-checkerboard AFM ordered but insulating.
AFM-$a$ is in fact not a truly AFM state, although we initially set the up- and down-spin moments the same in the amplitude in our calculation.
But after full optimization, we find that it is actually a ferrimagnetic state with a net magnetization about 1.97 $\mu_{\text{B}}$ per unit cell.

In $\beta$-Fe$_4$Se$_5$, the energy of the $\sqrt{5}\times\sqrt{5}$-pair-checkerboard AFM state is 129.9 meV/Fe lower than the $\sqrt{5}\times\sqrt{5}$-blocked-checkerboard AFM one.
In contrast, the energy of the former is about 95.8 meV/Fe higher than the latter in K$_2$Fe$_4$Se$_5$.
Apparently, the difference results from the intercalation of K atoms in Fe$_4$Se$_5$:
First, the intercalated K atoms impose a tensile strain on each Fe$_4$Se$_5$ layer, which enlarges the $a$-axis lattice constant from 8.41 {\AA} in $\beta$-Fe$_4$Se$_5$ to 8.69 {\AA} in K$_2$Fe$_4$Se$_5$.
Second, the intercalated atoms dope electrons into Fe$_4$Se$_5$ layers.

\begin{figure}[b]
\begin{center}
\includegraphics[width=8.6cm]{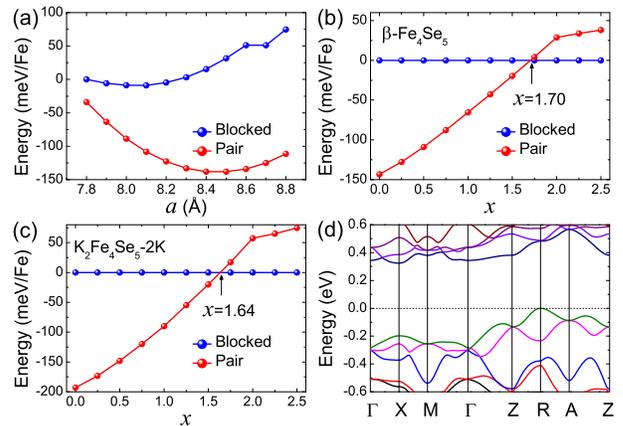}
  \caption{(Color online)
  Comparison between the $\sqrt{5} \times \sqrt{5}$-pair-checkerboard AFM and the $\sqrt{5} \times \sqrt{5}$-blocked-checkerboard AFM states.
  (a) Total energies of these two AFM orders in $\beta$-Fe$_4$Se$_5$ as functions of the $a$-axis lattice constant with tetragonal unit cell.
  The energy of the $\sqrt{5} \times \sqrt{5}$-blocked-checkerboard AFM state at $a$=7.8 {\AA} is set to zero.
  (b) Relative energy of the $\sqrt{5} \times \sqrt{5}$-pair-checkerboard AFM state (red) with respect to the $\sqrt{5} \times \sqrt{5}$-blocked-checkerboard AFM state (blue) as a function of doped electron concentration $x$ for $\beta$-Fe$_4$Se$_5$. (c) Same as in (b) but for K$_2$Fe$_4$Se$_5$-2K.
  (d) Band structure of the $\sqrt{5} \times \sqrt{5}$-blocked-checkerboard AFM state in the electron doped $\beta$-Fe$_4$Se$_5$ with $x$=2.}
\label{fig:Dope}
\end{center}
\end{figure}

To understand the effect of tensile strain introduced by K atoms, we calculate the energies of $\sqrt{5}\times\sqrt{5}$-pair-checkerboard and $\sqrt{5} \times \sqrt{5}$-blocked-checkerboard AFM states as a function of the $a$-axis lattice constant for $\beta$-Fe$_4$Se$_5$ with tetragonal unit cell.
The lattice constant along the $c$-axis and internal atomic positions are relaxed to minimize the total energy.
In a wide range of the $a$-axis lattice constant, as shown in Fig.~\ref{fig:Dope}(a), we find that the $\sqrt{5}\times\sqrt{5}$-pair-checkerboard AFM order is more stable than the $\sqrt{5}\times\sqrt{5}$-blocked-checkerboard AFM one.
It suggests that the difference between $\beta$-Fe$_4$Se$_5$ and K$_2$Fe$_4$Se$_5$ is not due to the tensile strain effect.

To understand the charge doping effect, we introduce extra electrons to $\beta$-Fe$_4$Se$_5$ and to K$_2$Fe$_4$Se$_5$ with all K atoms removed (namely, K$_2$Fe$_4$Se$_5$-2K).
A jellium background with positive charge is introduced to compensate the doped electrons and to avoid numerical divergence in the calculation.
Note that different AFM states have different optimized lattice constants.
For example, the optimized lattice constants are $a$=8.4682 {\AA} and $c$=6.5241 {\AA} in the $\sqrt{5} \times \sqrt{5}$-pair-checkerboard AFM state, and $a$=8.1092 {\AA} and $c$=6.2119 {\AA} in the $\sqrt{5}\times\sqrt{5}$-blocked-checkerboard AFM order of $\beta$-Fe$_4$Se$_5$.
In order to avoid the complication introduced by the lattice parameters, we do the calculation by taking the experimental lattice constants, i.e., $a$=8.41 {\AA} and $c$=5.47 {\AA} for $\beta$-Fe$_4$Se$_5$ \cite{Chen-PNAS}, $a$=8.6929 {\AA} and $c$=14.0168 {\AA} for K$_2$Fe$_4$Se$_5$-2K \cite{Bao-CPL28}.

For both materials without electron doping, we find that the $\sqrt{5} \times \sqrt{5}$-pair-checkerboard AFM state is lower in energy than the $\sqrt{5} \times \sqrt{5}$-blocked-checkerboard one [Figs.~\ref{fig:Dope}(b) and \ref{fig:Dope}(c)].
Upon electron doping, the energy difference between these two states gradually decreases.
A reversion in the total energy happens at a critical doping roughly equal to 1.7 electrons per formula for both $\beta$-Fe$_4$Se$_5$ and K$_2$Fe$_4$Se$_5$-2K.
The $\sqrt{5}\times\sqrt{5}$-blocked-checkerboard AFM state is 57.5 meV/Fe lower than the $\sqrt{5}\times\sqrt{5}$-pair-checkerboard one by adding 2 electrons per formula to K$_2$Fe$_4$Se$_5$-2K.
Here K$_2$Fe$_4$Se$_5$-2K+2$e$ is isoelectronic to K$_2$Fe$_4$Se$_5$.
Thus adding electron to Fe$_4$Se$_5$ layers reproduces correctly the ground-state AFM order of K$_2$Fe$_4$Se$_5$.
It suggests that the difference in the magnetic orders between $\beta$-Fe$_4$Se$_5$ and K$_2$Fe$_4$Se$_5$ results mainly from the doping effect introduced by the intercalated K atoms.
Furthermore, by doping 2 electrons per formula to $\beta$-Fe$_4$Se$_5$, we find that the metallic $\sqrt{5}\times\sqrt{5}$-blocked-checkerboard AFM state becomes insulating, with an energy gap of 324 meV [Fig.~\ref{fig:Dope}(d)].
This is reminiscent of the insulating AFM state of K$_2$Fe$_4$Se$_5$ \cite{Yan-PRB83}.

With the increase of electron doping, we find that the nearest-neighboring Fe-Fe bond length inside the inner Fe block decreases.
This leads to a tetramer lattice distortion of Fe atoms, in both the $\sqrt{5} \times \sqrt{5}$-blocked-checkerboard and the $\sqrt{5} \times \sqrt{5}$-pair-checkerboard AFM states.
The tetramer distortion also exists in the ground state of K$_2$Fe$_4$Se$_5$ \cite{Yan-PRB83}.
It  is in fact this tetramer distortion that stablizes the $\sqrt{5} \times \sqrt{5}$-blocked-checkerboard AFM order in K$_2$Fe$_4$Se$_5$.
To check how important this effect is in $\beta$-Fe$_4$Se$_5$, we carry out a total-energy calculation for this material without any tetramer distortion by using the lattice parameters obtained from the bulk $\beta$-FeSe.
We find that the $\sqrt{5}\times\sqrt{5}$-blocked-checkerboard AFM state is about 27.8 meV/Fe higher in energy than the $\sqrt{5}\times\sqrt{5}$-pair-checkerboard one
at the doping of 2 electrons per Fe$_4$Se$_5$ formula.
This shows that it is the tetramer lattice distortion enhanced by electron doping that lowers the energy of the $\sqrt{5}\times\sqrt{5}$-blocked-checkerboard AFM state.

The above results are obtained without considering the van der Waals (vdW) interaction between FeSe layers.
This overestimates the lattice constant along the $c$-axis.
If the vdW correction\cite{Grimme-JCC,Barone-JCC} is taken into account, we find that the ground state is still $\sqrt{5} \times \sqrt{5}$-pair-checkerboard AFM insulator.
The band gap is 172 meV, and the $a$- and $c$-axis lattice constants are respectively 8.2440 {\AA} and 5.5619 {\AA}, in agreement with the experimental results \cite{Chen-PNAS}.

\section{SUMMARY}

In conclusion, we have studied the electronic and magnetic structures of $\beta$-Fe$_4$Se$_5$.
Similar to K$_2$Fe$_4$Se$_5$, we find that the ground state of $\beta$-Fe$_4$Se$_5$ is an AFM insulator.
But the ground-state is $\sqrt{5}\times\sqrt{5}$-pair-checkerboard AFM ordered, unlike in K$_2$Fe$_4$Se$_5$ where the ground state is $\sqrt{5}\times\sqrt{5}$-blocked-checkerboard AFM ordered.
The band excitation gap of $\beta$-Fe$_4$Se$_5$ is 290 meV, in agreement with the experimental result.
The $\sqrt{5}\times\sqrt{5}$-pair-checkerboard AFM order is robust against the lattice  tensile strain.
But electron doping, introduced by extra Fe or intercalated alkali-metal atoms, can turn it into a $\sqrt{5}\times\sqrt{5}$-blocked-checkerboard AFM state above a critical doping level.

\begin{acknowledgments}
This work is supported by National Natural Science Foundation of China (Grant Nos. 11404383, 91421304, 11474331, and 11474004). M.G. is also supported by Zhejiang Provincial
Natural Science Foundation of China under Grant No. LY17A040005 and K.C.Wong Magna Fund in Ningbo University.
\end{acknowledgments}


\begin{references}

\bibitem{Kamihara-JACS}Y. Kamihara, T. Watanabe, M. Hirano, and H. Hosono,
\textit{Iron-Based Layered Superconductor La[O$_{1-x}$F$_x$]FeAs ($x$=0.05-0.12) with $T_c$=26 K},
J. Am. Chem. Soc. {\bf 130}, 3296 (2008).


\bibitem{Zhao-NatMater7}
J. Zhao, Q. Huang, C. de la Cruz, S. Li, J. W. Lynn, Y. Chen, M. A. Green, G. F. Chen, G. Li, Z. Li, J. L. Luo, N. L. Wang, and P. Dai,
\emph{Structural and magnetic phase diagram of CeFeAsO$_{1-x}$F$_x$ and its relation to high-temperature superconductivity},
\href{http://dx.doi.org/10.1038/nmat2315}{Nat. Mater. {\bf7}, 953 (2008).}

\bibitem{Okabe-PRB}H. Okabe, N. Takeshita, K. Horigane, T. Muranaka, and J. Akimitsu,
\emph{Pressure-induced high-$T_c$ superconducting phase in FeSe: Correlation between anion height and $T_c$},
\href{http://dx.doi.org/10.1103/PhysRevB.81.205119}{Phys. Rev. B {\bf 81}, 205119 (2010).}

\bibitem{Mizuguchi-SST}Y. Mizuguchi, Y. Hara, K. Deguchi, S. Tsuda, T. Yamaguchi, K. Takeda, H. Kotegawa, H. Tou, and Y. Takano,
\emph{Anion height dependence of $T_c$ for the Fe-based superconductor},
\href{http://dx.doi.org/10.1088/0953-2048/23/5/054013}{Supercond. Sci. Technol. {\bf 23}, 054013 (2010).}

\bibitem{Hsu-PNAS}
F.-C. Hsu, J.-Y. Luo, K.-W. Yeh, T.-K. Chen, T.-W. Huang, P. M. Wu, Y.-C. Lee, Y.-L. Huang, Y.-Y. Chu, D.-C. Yan, and M.-K. Wu,
\emph{Superconductivity in the PbO-type structure $\alpha$-FeSe},
Proc. Natl. Acad. Sci. U.S.A. {\bf 105}, 14262 (2008).

\bibitem{Guo-PRB82}J. Guo, S. Jin, G. Wang, S. Wang, K. Zhu, T. Zhou, M. He, and X. Chen,
\emph{Superconductivity in the iron selenide K$_x$Fe$_2$Se$_2$ (0$\leq$$x$$\leq$1.0)},
Phys. Rev. B {\bf 82}, 180520(R) (2010).

\bibitem{Maziopa-JPCM}A. Krzton-Maziopa, Z. Shermadini, E. Pomjakushina, V. Pomjakushin, M. Bendele, A. Amato, R. Khasanov, H. Luetkens, and K. Conder,
\emph{Synthesis and crystal growth of Cs$_{0.8}$(FeSe$_{0.98}$)$_2$: a new iron-based superconductor with $T_c$ = 27 K},
\href{http://dx.doi.org/10.1088/0953-8984/23/5/052203}{J. Phys. Condens. Matter {\bf 23},052203 (2011).}

\bibitem{Fang-EPL}M.-H. Fang, H.-D. Wang, C.-H. Dong, Z.-J. Li, C.-M. Feng, J. Chen, and H. Q. Yuan,
\emph{Fe-based superconductivity with $T_c$=31K bordering an antiferromagnetic insulator in (Tl,K)Fe$_x$Se$_2$},
\href{http://dx.doi.org/10.1209/0295-5075/94/27009}{Europhys. Lett. {\bf 94}, 27009 (2011).}

\bibitem{Wang-EPL93}H.-D. Wang, C.-H. Dong, Z.-J. Li, Q.-H. Mao, S.-S. Zhu, C.-M. Feng, H. Q. Yuan, and M.-H. Fang,
\emph{Superconductivity at 32 K and anisotropy in Tl$_{0.58}$Rb$_{0.42}$Fe$_{1.72}$Se$_2$ crystals},
\href{http://dx.doi.org/10.1209/0295-5075/93/47004}{Europhys. Lett. {\bf 93}, 47004 (2011).}

\bibitem{Sun-Nature}
L. Sun, X.-J. Chen, J. Guo, P. Gao, Q.-Z. Huang, H. Wang, M. Fang, X. Chen, G. Chen, Q. Wu, C. Zhang, D. Gu, X. Dong, L. Wang, K. Yang, A. Li, X. Dai, H.-k. Mao, and Z. Zhao,
\emph{Re-emerging superconductivity at 48 kelvin in iron chalcogenides},
\href{http://dx.doi.org/10.1038/nature10813}{Nature {\bf 483}, 67 (2012).}

\bibitem{Ying-Sci2}T. P. Ying, X. L. Chen, G. Wang, S. F. Jin, T. T. Zhou, X. F. Lai, H. Zhang and W. Y. Wang,
\emph{Observation of superconductivity at 30$\sim$46K in A$_x$Fe$_2$Se$_2$ (A=Li, Na, Ba, Sr, Ca, Yb, and Eu)},
\href{http://dx.doi.org/10.1038/srep00426}{Sci. Rep. {\bf 2}, 426 (2012).}

\bibitem{Yan-PRL}X.-W. Yan, M. Gao, Z.-Y. Lu, and T. Xiang,
\emph{Electronic Structures and Magnetic Order of Ordered-Fe-Vacancy Ternary Iron Selenides TlFe$_{1.5}$Se$_2$ and $A$Fe$_{1.5}$Se$_2$
($A$=K, Rb, or Cs)},
\href{http://dx.doi.org/10.1103/PhysRevLett.106.087005}{Phys. Rev. Lett. {\bf 106}, 087005 (2011).}

\bibitem{Yan-PRB83}X.-W. Yan, M. Gao, Z.-Y. Lu, and T. Xiang,
\emph{Ternary iron selenide K$_{0.8}$Fe$_{1.6}$Se$_2$ is an antiferromagnetic semiconductor},
\href{http://dx.doi.org/10.1103/PhysRevB.83.233205}{Phys. Rev. B {\bf 83}, 233205 (2011).}

\bibitem{Bao-CPL28}W. Bao, Q. Huang, G. F. Chen, M. A. Green, D. M. Wang, J. B. He, X. Q. Wang, and Y. Qiu,
\emph{A Novel Large Moment Antiferromagnetic Order in K$_{0.8}$Fe$_{1.6}$Se$_2$ Superconductor},
\href{http://dx.doi.org/10.1088/0256-307X/28/8/086104}{Chin. Phys. Lett. {\bf 28}, 086104 (2011).}

\bibitem{Zhao-PRL109}J. Zhao, H. Cao, E. Bourret-Courchesne, D.-H. Lee, and R. J. Birgeneau,
\emph{Neutron-Diffraction Measurements of an Antiferromagnetic Semiconducting Phase in the Vicinity of the High-Temperature Superconducting State of K$_x$Fe$_{2-y}$Se$_2$},
\href{http://dx.doi.org/10.1103/PhysRevLett.109.267003}{Phys. Rev. Lett. {\bf 109}, 267003 (2012).}

\bibitem{Chen-PRX1}
F. Chen, M. Xu, Q. Q. Ge, Y. Zhang, Z. R. Ye, L. X. Yang, Juan Jiang, B. P. Xie, R. C. Che, M. Zhang, A. F. Wang, X. H. Chen, D. W. Shen, J. P. Hu, and D. L. Feng,
\emph{Electronic Identification of the Parental Phases and Mesoscopic Phase Separation of K$_x$Fe$_{2-y}$Se$_2$ Superconductors},
\href{http://dx.doi.org/10.1103/PhysRevX.1.021020}{Phys. Rev. X {\bf 1}, 021020 (2011).}

\bibitem{Ding-Natcomm4}X. Ding, D. Fang, Z. Wang, H. Yang, J. Liu, Q. Deng,	G. Ma, C. Meng, Y. Hu, and H.-H. Wen,
\emph{Influence of microstructure on superconductivity in K$_x$Fe$_{2-y}$Se$_2$ and evidence for a new parent phase K$_2$Fe$_7$Se$_8$},
Nat. Commun. {\bf 4}, 1897 (2013).

\bibitem{Pomjakushin-JPCM}V. Y. Pomjakushin, A. Krzton-Maziopa, E. V. Pomjakushina, K. Conder, D. Chernyshov, V. Svitlyk, and A. Bosak,
\emph{Intrinsic crystal phase separation in the antiferromagnetic superconductor Rb$_y$Fe$_{2-x}$Se$_2$: a diffraction study},
\href{http://dx.doi.org/10.1088/0953-8984/24/43/435701}{J. Phys.: Condens. Matter {\bf 24}, 435701 (2012).}

\bibitem{Ricci-PRB84}
A. Ricci, N. Poccia, G. Campi, B. Joseph, G. Arrighetti, L. Barba, M. Reynolds, M. Burghammer, H. Takeya, Y. Mizuguchi, Y. Takano, M. Colapietro, N. L. Saini, and A. Bianconi,
\emph{Nanoscale phase separation in the iron chalcogenide superconductor K$_{0.8}$Fe$_{1.6}$Se$_2$ as seen via scanning nanofocused x-ray diffraction},
\href{http://dx.doi.org/10.1103/PhysRevB.84.060511}{Phys. Rev. B {\bf 84}, 060511(R) (2011).}

\bibitem{Ricci-PRB91}
A. Ricci, N. Poccia, B. Joseph, D. Innocenti, G. Campi, A. Zozulya, F. Westermeier, A. Schavkan, F. Coneri, A. Bianconi, H. Takeya, Y. Mizuguchi, Y. Takano, T. Mizokawa, M. Sprung, and N. L. Saini,
\emph{Direct observation of nanoscale interface phase in the superconducting chalcogenide K$_x$Fe$_{2-y}$Se$_2$ with intrinsic phase separation},
\href{http://dx.doi.org/10.1103/PhysRevB.91.020503}{Phys. Rev. B {\bf 91}, 020503(R) (2015).}

\bibitem{Wang-PRB83}Z. Wang, Y. J. Song, H. L. Shi, Z. W. Wang, Z. Chen, H. F. Tian, G. F. Chen, J. G. Guo, H. X. Yang, and J. Q. Li,
\emph{Microstructure and ordering of iron vacancies in the superconductor system K$_y$Fe$_x$Se$_2$ as seen via transmission electron microscopy},
\href{http://dx.doi.org/10.1103/PhysRevB.83.140505}{Phys. Rev. B {\bf 83}, 140505 (2011).}

\bibitem{Li-PRL109}
W. Li, H. Ding, Z. Li, P. Deng, K. Chang, K. He, S. Ji, L. Wang, X. Ma, J.-P. Hu, X. Chen, and Q.-K. Xue,
\emph{KFe$_2$Se$_2$ is the Parent Compound of K-Doped Iron Selenide Superconductors},
\href{http://dx.doi.org/10.1103/PhysRevLett.109.057003}{Phys. Rev. Lett. {\bf 109}, 057003 (2012).}

\bibitem{Zhang-PRB} G. M. Zhang, Z. Y. Lu, T. Xiang,
\emph{Superconductivity mediated by the antiferromagnetic spin-wave in chalcogenide iron-base superconductors},
Phys. Rev. B {\bf 84}, 052502 (2011).

\bibitem{Chen-PNAS}
T.-K. Chen, C.-C. Chang, H.-H. Chang, A.-H. Fang, C.-H. Wang, W.-H. Chao, C.-M. Tseng, Y.-C. Lee, Y.-R. Wu, M.-H. Wen, H.-Y. Tang, F.-R. Chen, M.-J. Wang, M.-K. Wu, and D. V. Dyck,
\emph{Fe-vacancy order and superconductivity in tetragonal $\beta$-Fe$_{1-x}$Se},
\href{http://dx.doi.org/10.1073/pnas.1321160111}{Proc. Natl. Acad. Sci. USA {\bf 111}, 63 (2014).}

\bibitem{Fang-PRB93}
Y. Fang, D. H. Xie, W. Zhang, F. Chen, W. Feng, B. P. Xie, D. L. Feng, X. C. Lai, and S. Y. Tan,
\emph{Tunable Fe-vacancy disorder-order transition in FeSe thin films},
\href{http://dx.doi.org/10.1103/PhysRevB.93.184503}{Phys. Rev. B {\bf 93}, 184503 (2016).}

\bibitem{pwscf}P. Giannozzi \emph{et al.},
\emph{QUANTUM ESPRESSO: a modular and open-source software project for quantum simulations of materials},
J. Phys.: Condens. Matter {\bf 21}, 395502 (2009).

\bibitem{pbe}J. P. Perdew, K. Burke, and M. Ernzerhof,
\emph{Generalized Gradient Approximation Made Simple},
Phys. Rev. Lett. {\bf 77}, 3865 (1996).

\bibitem{vanderbilt}D. Vanderbilt,
\emph{Soft self-consistent pseudopotentials in a generalized eigenvalue formalism},
Phys. Rev. B {\bf 41}, 7892 (1990).

\bibitem{Sabrowsky-JMMM}H. Sabrowsky, M. Rosenberg, D. Welz, P. Deppe, and W. Sch\"{a}fer,
\emph{A neutron and M\"{o}ssbauer study of TlFe$_x$S$_2$ compounds},
\href{http://dx.doi.org/10.1016/0304-8853(86)90900-5}{J. Magn. Magn. Mater. {\bf 54-57}, 1497 (1986).}


\bibitem{Cao-arXiv1407}H.-Y. Cao, S. Chen, H. Xiang, X.-G. Gong,
\emph{Antiferromagnetic ground state with pair-checkerboard order in FeSe},
\href{http://dx.doi.org/10.1103/PhysRevB.91.020504}{Phys. Rev. B {\bf 91}, 020504(R) (2015).}

\bibitem{Cruz-Nature}
C. de la Cruz, Q. Huang, J. W. Lynn, J. Li, W. Ratcliff II, J. L. Zarestky, H. A. Mook, G. F. Chen, J. L. Luo, N. L. Wang, P. Dai,
\emph{Magnetic order close to superconductivity in the iron-based layered LaO$_{1-x}$F$_x$FeAs systems},
\href{http://dx.doi.org/10.1038/nature07057}{Nature {\bf 453}, 899 (2008).}

\bibitem{Rotter-PRB78}M. Rotter, M. Tegel, D. Johrendt, I. Schellenberg, W. Hermes, and R. P\"{o}ttgen,
\emph{Spin-density-wave anomaly at 140 K in the ternary iron arsenide BaFe$_2$As$_2$},
\href{http://dx.doi.org/10.1103/PhysRevB.78.020503}{Phys. Rev. B {\bf 78}, 020503(R) (2008).}

\bibitem{McQueen-PRL103}T. M. McQueen, A. J. Williams, P. W. Stephens, J. Tao, Y. Zhu, V. Ksenofontov, F. Casper, C. Felser, and R. J. Cava,
\emph{Tetragonal-to-Orthorhombic Structural Phase Transition at 90 K in the Superconductor Fe$_{1.01}$Se},
\href{http://dx.doi.org/10.1103/PhysRevLett.103.057002}{Phys. Rev. Lett {\bf 103}, 057002 (2009).}

\bibitem{Ma-PRL102}F. Ma, W. Ji, J. Hu, Z.-Y. Lu, and T. Xiang,
\emph{First-Principles Calculations of the Electronic Structure of Tetragonal $\alpha$-FeTe
and $\alpha$-FeSe Crystals: Evidence for a Bicollinear Antiferromagnetic Order},
Phys. Rev. Lett. {\bf 102}, 177003 (2009).

\bibitem{Grimme-JCC}S. Grimme,
\emph{Semiempirical GGA-type density functional constructed with a long-range dispersion correction},
\href{http://dx.doi.org/10.1002/jcc.20495}{J. Comp. Chem. {\bf 27}, 1787 (2006).}

\bibitem{Barone-JCC}V. Barone, M. Casarin, D. Forrer, M. Pavone, M. Sambi, and A. Vittadini,
\emph{Role and effective treatment of dispersive forces in materials: Polyethylene and graphite crystals as test cases},
\href{http://dx.doi.org/10.1002/jcc.21112}{J. Comp. Chem. {\bf 30}, 934 (2009).}

\end{references}
\end{document}